# On the interpretation of clustering from the angular APM Galaxy Survey

Enrique Gaztañaga
*Centre d'Estudis Avançats de Blanes, C.S.I.C., Cami de Santa Barbara, 17300 Blanes, Girona, Spain*
*Department of Physics, Astrophysics, University of Oxford, Keble Road, Oxford OX1 3RH*

**ABSTRACT**

We analyze the uncertainties in the amplitudes of the spatial correlation functions estimated from angular correlations in a sample from the APM Galaxy Survey, with $b_J = 17 - 20$. We model the uncertainties in the selection function and in the evolution of clustering. In particular we estimate $\sigma_8^{APM}$, the rms galaxy number fluctuations in spheres of radius at $8\,h^{-1}$ Mpc, from the measured angular variance in the APM. The uncertainty in $\sigma_8^{APM}$ has three main contributions: 8% from sampling and selection function uncertainties, 7% from the uncertainty in the evolution of clustering and 3% from the uncertainty in the value of $\Omega_0$. Including all these contributions, we find $\sigma_8^{APM}$ is in the range $0.78 - 1.08$. If the galaxy clustering in the APM evolves as expected from gravitational clustering of matter fluctuations, then $\sigma_8^{APM} = 0.95 \pm 0.07$ ($1.00 \pm 0.08$) for $\Omega_0 \simeq 1$ ($\Omega_0 \simeq 0$), close to the values for nearby optical samples. On the other hand, if we assume that clustering evolution is fixed in comoving coordinates $\sigma_8^{APM} = 0.83 \pm 0.05$ ($0.87 \pm 0.06$), closer to the results for nearby IRAS samples. The final uncertainty in the range of values for the hierarchical amplitudes $S_J \equiv \overline{\xi}_J/\overline{\xi}_2^{J-1}$ is typically twice the estimated sampling errors, with the highest values for the case of less clustering evolution. We compare our estimates with other results and discuss the implications for models of structure formation.

## 1 INTRODUCTION

Angular catalogs of galaxy positions have proven a very useful tool with which to study the statistical properties of large scale density fluctuations, as they provide large volume coverage, so that local density fluctuations in the nearby galaxy distribution are averaged out. The isotropy and large scale homogeneity of the universe allows the recovery of the underlying spatial statistics. However, in the derivation of three-dimensional (3D) properties from angular data there are large uncertainties which arise from the selection function and the evolution of the galaxy population. These uncertainties must be considered carefully for an accurate interpretation of the underlying clustering properties.

We concentrate on the APM Galaxy Survey (Maddox et al. 1990a-c) of angular positions and, in particular, on the $b_J = 17 - 20$ subsample which has over $1.3 \times 10^6$ galaxies and a mean depth $\mathcal{D} \sim 400\,h^{-1}$ Mpc. Maddox et al. (1990a) have proposed a redshift dependent luminosity function model as a rough approximation to the selection function used to estimate the 3D clustering in the APM. Here we use new observational constraints on the local luminosity function (Loveday et al. 1992) to study the range of possible selection functions and their effect on the estimated 3D clustering properties, also extending the analysis to higher order correlations.

We will pay special attention to the normalization of clustering amplitudes $\overline{\xi}_J$ at small scales, where power-laws are a good approximation. As the APM covers the largest volume sampled to date, well over $1.5 \times 10^8\,(h^{-1}\,\mathrm{Mpc})^3$, an accurate determination of these quantities should provide important constraints on models of structure formation.

The projection effects and the recovery of 3-D clustering are presented in § 2. In § 3 we discuss limits on the selection function of our sample by modelling the luminosity function, while in § 4 we consider models for the clustering evolution. We apply these models to the problem of inversion from the APM data in § 5 and present a final discussion in § 6.

## 2 PROJECTION EFFECTS

We now present a simple method for recovering the 3-D variance, $\overline{\xi}_2(R)$, and higher order moments, $\overline{\xi}_J(R)$, from the 2-D correlations, $\overline{\omega}_J(\theta)$. We will use the same notation and definitions as in Gaztañaga (1994, hereafter G94).

## 2.1 Scale-invariant model

Consider first the following scale-invariant model for the correlations:

$$\xi_2(r) = \left(\frac{r_0}{r}\right)^\gamma, \quad (1)$$

$$\xi_J(r_1,\ldots,r_J) = Q_J \sum_{ab} \prod^{J-1} \xi_2(r_{ab}), \quad J > 2 \quad (2)$$

(e.g. Fry & Peebles 1978, Fry 1984b). Here the product of the two-point functions, $\xi_2(r_{ij})$, is over $J-1$ independent pairs of relative separations and the sum, consisting of $J^{J-2}$ terms, is over equivalent reassignments of labels $i,j = 1, 2, 3\ldots, J$. The amplitudes $Q_J$ are just numbers that can be generalized to $Q_{J,\alpha}$ where $\alpha$ denotes different topologies in the graphs connecting the labels. Thus, the hierarchy in equation (2) is composed of "tree" graphs (connected with no cycles) of $J$ vertices and $J-1$ edges.

Observations indicate that the above hierarchy holds at least for lower values of $J$, at small scales (Groth & Peebles 1977; Fry & Peebles 1978) or up to $J = 10$ when averaged over all scales (Szapudi, Szalay & Boshan 1992; Meiksin, Szapudi & Szalay 1992, Szapudi et al. 1995). This same hierarchy has also been obtained from theoretical considerations (Davis & Peebles 1977, Fry 1984a, Hamilton 1988, Balian & Schaeffer 1989a,1989b). In perturbation theory (PT), similar hierarchical forms to equation (2) have been found (Peebles 1980, Fry 1984b, Goroff et al. 1986, Bernardeau 1994). In this case, the hierarchical parameters $Q_J$ are not constant, but depend on the configuration arrangements and, in particular, on the scale. Analyses of N-body simulations show that these analytical results are accurate on large scales for smoothed correlations $\overline{\xi}_J$ (Juszkiewicz, Bouchet & Colombi 1993, Bernardeau 1994, Juszkiewicz et al. 1995, Lokas et al. 1995, Gaztañaga & Baugh 1995, Baugh, Gaztañaga & Efstathiou 1995).

The above model produces the following volume averaged correlations in a spherical cell (top-hat window smoothing) of radius $R$:

$$\overline{\xi}_2(R) = \sigma_8^2 \left(\frac{8 h^{-1} \text{Mpc}}{R}\right)^\gamma \quad (3)$$

$$\overline{\xi}_J(R) = S_J [\overline{\xi}_2(R)]^{J-1}, \quad J > 2$$

The smoothed amplitudes, $\sigma_8$ and $S_J$ are related to the multi-point amplitudes, $r_0$ and $Q_J$, by:

$$\sigma_8^2 = \frac{2^{-\gamma}}{(1-\gamma/3)(1-\gamma/4)(1-\gamma/6)} \left(\frac{r_0}{8 h^{-1} \text{Mpc}}\right)^\gamma, \quad (4)$$

$$S_J = B_J J^{J-2} Q_J, \quad (5)$$

where $B_J \sim 1$ are given in G94 (see also Boschan, Szapudi & Szalay 1994).

For small angles this scale invariant model produces the following angular averaged correlations:

$$\overline{\omega}_2(\theta) = A \, \theta^{1-\gamma} \quad (6)$$

$$\overline{\omega}_J(\theta) = s_J [\overline{\omega}_2(\theta)]^{J-1}, \quad J > 2$$

The angular amplitudes, $A$ and $s_J$ are related to the multi-point values, $r_0$ and $Q_J$, by:

$$A = r_0^\gamma \, T_\gamma \, \frac{I_2}{I_1^2} \, \frac{\Gamma(1/2)\,\Gamma(\gamma-1/2)}{\Gamma(\gamma/2)}$$

$$s_J = r_J \, S_J \, \left(\frac{C_J}{B_J}\right)$$

$$r_J \equiv \frac{I_1^{J-2} I_J}{I_2^{J-1}} \quad (7)$$

where $C_J \simeq 1$ are given in G94 (see also Boschan, Szapudi & Szalay 1994) and $T_\gamma$ is a geometrical factor that comes from the area average in $\overline{\omega}_2$:

$$T_\gamma = \frac{4}{\pi(5-\gamma)} \int_0^1 x \, dx \int_0^{2\pi} d\phi \left(1 + x^2 - 2x \cos\phi\right)^{\frac{1-\gamma}{2}}. \quad (8)$$

Note that the $r_J$ in equation (7) are dimensionless and are not directly related to $r_0$ (which has units of $h^{-1}$ Mpc). The values of $I_k$ in equation (7), can be expressed as:

$$I_k = \int_0^\infty F(x) \, x^2 dx \, \psi^k \, x^{(3-\gamma)(k-1)} \, (1+z)^{(3+\epsilon-\gamma)(1-k)} \quad (9)$$

where $x$ is the comoving coordinate. These integrals depend on the selection function $\psi(x)$, the curvature correction $F^2(x) = [1 - (\Omega_0 - 1)(H_0 x/c)^2]$ and the evolution of clustering, parametrized by $\epsilon$ (which will be discussed in the next sections).

Thus for a given scale-invariant model with slope $\gamma$, it is possible to use the above expressions to relate the estimated angular amplitudes $A$ and $s_J$, in equation (7), to the underlying three dimensional amplitudes, i.e. $\sigma_8$ and $S_J$, from equations 4 and 5.

## 2.2 Quasi-scale-invariant model

Consider now a distribution that is not exactly scale-invariant but has correlations $\xi_J$ that can be parametrized as a scale-invariant distribution as in equation (2) with $r_0$, $\gamma$ and $Q_J$ being a slowly varying function of scale. We call this a quasi-scale-invariant model.

For a a quasi-scale-invariant model it should be possible to apply a local inversion at each scale. In principle the correlations on all scales $R$ contribute to the correlations on angular scale $\theta$, but because the sample has a finite depth, $\mathcal{D}$, there is a characteristic scale $R \simeq \mathcal{D}\theta$. In our analysis we relate angular scales $\theta$ to 3-D scales using $R = \mathcal{D}\theta$, where $\mathcal{D}$ is the estimated distance which corresponds to the mean redshift of the sample (see also Peebles 1980). Although there is some ambiguity as to what the best definition of $\mathcal{D}$ should be, in the scale-invariant regime, we find that the estimated amplitudes of $\overline{\xi}_J$ are insensitive to changes in our chosen value of $\mathcal{D}$.

Thus at a given scale $\theta$ with local slope $\gamma$, we use the above expressions to relate the estimated local angular amplitudes $A$ and $s_J$ to the underlying three dimensional values, i.e. $\sigma_8$ and $S_J$ at scale $R = \mathcal{D}\theta$. This model was used in G94 to recover the 3D correlations in the APM Survey.

## 2.3 Tests on N-body simulations

We have tested the quasi-scale invariant method on several simulations with different shapes and amplitudes for the variance $\bar{\xi}_2$. We have tried both galaxy and cluster simulations, generated by Gaztañaga & Baugh (1995) and Croft & Efstathiou (1994a,b). We first estimate $\bar{\xi}_J$ using the counts in cells method for the whole simulation box (as in Baugh, Gaztañaga & Efstathiou 1995). Next we transform the simulation into an observational catalogue with a given selection function. ¿From this mock catalogue we estimate $\bar{w}_J$ and use the above method to recover $\bar{\xi}_J$. The comparison (Baugh & Gaztañaga, in preparation) shows excellent agreement within the errors even at large scales, where there is a significant break from the power-law model.

The simulations we use have values of $S_J$ which show a small variation with scale, e.g. $S_3 \simeq R^\alpha$, with $\alpha \simeq 0.1$ (Fig. 3 in Gaztañaga & Baugh 1995), and strictly speaking neither the scale-invariant nor the quasi-scale-invariant models should be used, as $S_J$ should be constants. Nevertheless, we still find reasonable agreement from the inversion when we compare local values of $S_J$. These results and further details are presented elsewhere.

## 2.4 Test on the APM galaxy data

We now apply this method to the APM galaxies and compare it with a the results of a previous estimate by Baugh & Efstathiou (1993). We first estimate the shape of the variance, $\bar{\xi}_2$, in the APM Galaxy Survey by integrating the three-dimensional P(k) measured by Baugh & Efstathiou (1994), i.e.

$$\bar{\xi}_2 = \frac{V}{2\pi^2} \int_{k_1}^{k_2} dk\, k^2 P(k) W^2(kR), \qquad (10)$$

where $W(kR)$ is the Fourier transform of the spherical window with radius $R$

$$W(kR) = \frac{3}{(kR)^3}[\sin(kR) - kR\cos(kR)]. \qquad (11)$$

The errors correspond to $2\sigma$ scatter in the angular two-point correlation function (Maddox et al. 1990a). The results are shown as filled symbols in Figure 1.

The second estimate is based on the method presented above for quasi-scale-invariant models. We use the angular variance measured in the APM (i.e. Figure 5). Our final errors include the sampling errors in $\bar{w}_2$ and the uncertainty in the local slope. In both cases we use the selection function of Baugh & Efstathiou (1993) and the same model for the evolution of clustering ($\epsilon = 0$, see below). Open squares with error bars in Figure 1 shows this new estimate of $\bar{\xi}_2$ compared to that from $P(k)$. Both results agree perfectly within the errors.

## 3 THE SELECTION FUNCTION

The selection function $\psi(x)$ is the normalized probability that a galaxy at coordinate $x$ is included in the catalogue. This probability is proportional to the estimated number of galaxies at this coordinate:

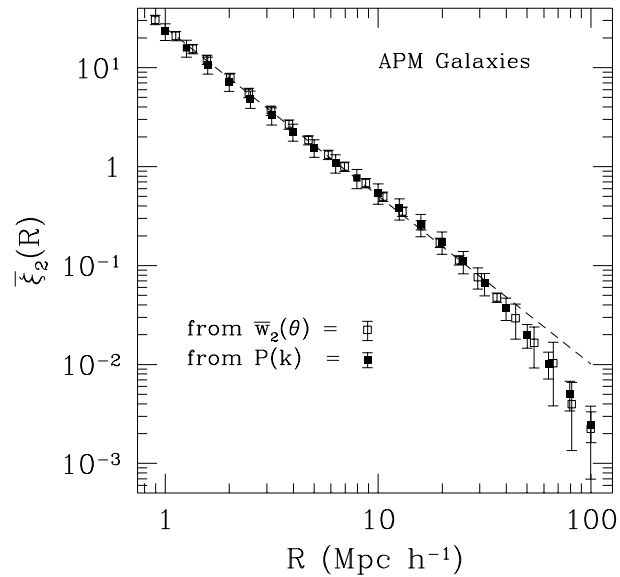

**Figure 1.** Comparison of two different estimates of $\bar{\xi}_2$ from the angular APM sample of galaxies. Filled squares are obtained by integrating the power spectrum $P(k)$ of Baugh & Efstathiou (1993). Open squares correspond to a direct inversion of the angular variance $\bar{w}_2(\theta)$.

$$\psi(x) = \psi^* \int_{q_1(x)}^{q_2(x)} dq\, \phi(q) \qquad (12)$$

where $\psi^*$ is adjusted so that the probability integrates to unity over the sample. $\phi(q)$ is the luminosity function and $q_1(x)$ and $q_2(x)$ are the scaled luminosities corresponding to the lower and upper limits in the range of apparent magnitudes used to build the galaxy sample or catalog under study. In our case these are $b_J = 17$ and $b_J = 20$ respectively. The luminosity function $\phi(q)$ in equation (12) should include all photometric contamination that has not been corrected for in the magnitudes in our sample, i.e. $\phi(q)$ should not have been corrected for magnitude errors, k-corrections or extinction. This is done in practice by introducing an *observer* magnitude system $M'$ so that the absolute magnitude, $M$, of a galaxy at redshift $z$ is given by $M' = M - kz$. We use $k = 3$ as a rough approximation but it should not affect our results much because we allow all our parameters have a redshift dependence. We also use the standard Schechter form for the luminosity function:

$$\phi(q) = \phi^* q^\alpha e^{-q}, \quad q = 10^{-\frac{2}{5}(M'-M^*)} \qquad (13)$$

An alternative to this approach of estimating $\psi(x)$ is to have a direct measurement of the redshift distribution $N(z)\,dz$, i.e. the number of galaxies at a given redshift $z$ (as in Baugh & Efstathiou 1993). However, a redshift catalog of APM galaxies in the range $b_J = 17 - 20$ is not yet available, and instead we will have to extrapolate $\psi(x)$ from luminosity function estimations.

## 3.1 Constraints on the local luminosity function

In the analysis of a bright sample of the APM galaxies (with $b_J < 17.15$), the Stromlo-APM Redshift Survey, Loveday et al. (1992) found that the values of the parameters in the luminosity function (uncorrected for magnitude errors) are: $\alpha_0 = -1.11 \pm 0.15$, $M_0^* = -19.73 \pm 0.13$ and $\phi_0^* = 1.12 \pm 0.12 \times 10^{-2} h^3$ Mpc$^{-3}$, where the errors in $\phi^*$ are basically due to the uncertainties in $\alpha$ and $M^*$. The median redshift of the Stromlo-APM sample is only $z \sim 0.05$ but the volume is large enough to avoid clustering effects or large scale density fluctuations. Similar results have recently been obtained by Vettolani et al. (1995) over a sample from the ESO key-project. If we apply these parameters for $\phi(q)$ to the $b_J = 17 - 20$ APM sample we find that it predicts a total number of $\sim 740,000$ galaxies. This is almost a factor of two smaller than the actual measured number, $\simeq 1,300,000$ galaxies. This seems to support the conclusion of Maddox et al. (1990b) that the rapid increase in number counts in the magnitude range $b_J = 16 - 19$ can only be explained by significant evolution of the galaxy population at redshifts $z \sim 0.1$. This could be a consequence of luminosity evolution or a decrease in the number density (see Koo and Kron 1992 and Colles 1994, Glazebrook et al. 1995a, 1995b for a recent discussion), but the observations could also be affected by important selection effects (e.g. Salzer 1994, McGaugh 1994). Koo, Gronwall & Bruzual (1993) and Gronwall & Koo (1995) have proposed that traditional luminosity evolution could explain the observations if we assume a particular (non Schechter) local luminosity function, with more galaxies (than in Loveday et al. extrapolation) at the faint end, $M_B > -16.5$, where there are no direct observations.

Whatever the reason, we will consider a rapid change in the *effective* luminosity function to parametrize the redshift distribution $N(z)$ in our magnitude range. As pointed out above, this approach might not be the most general possibility, but it seems that it can reproduce the observed $N(z)$ distributions in the cases where data is available.

## 3.2 Constraints on evolution with redshift

We have parametrized the redshift evolution by making $\phi^* = \phi^*(z)$, $M^* = M^*(z)$ and $\alpha = \alpha(z)$ function of the redshift $z$. Because the redshifts are small $z < 1$ we use linear functions:

$$\begin{aligned} M^* &= M_0^* + M_1^* \, z \, , \\ \alpha &= \alpha_0 + \alpha_1 \, z \, , \\ \phi^* &= \phi_0^* \, (1 + \phi_1^* \, z) \, , \end{aligned} \qquad (14)$$

so that the dimensionless parameters $M_1^*$, $\alpha_1$ and $\phi_1^*$ measure the redshift dependence. In order to set some constraints on the parameter space we use the following results:

1) The Stromlo-APM values of the luminosity function at low redshifts from Loveday et al. (1992) are taken as corresponding to $z \simeq 0$. Therefore, we take:

$$\begin{aligned} \alpha_0 &= -1.11 \pm 0.15 \\ M_0^* &= -19.73 \pm 0.13 \\ \phi_0^* &= (1.12 \pm 0.12) \times 10^{-2} h^3 Mpc^{-3} \end{aligned} \qquad (15)$$

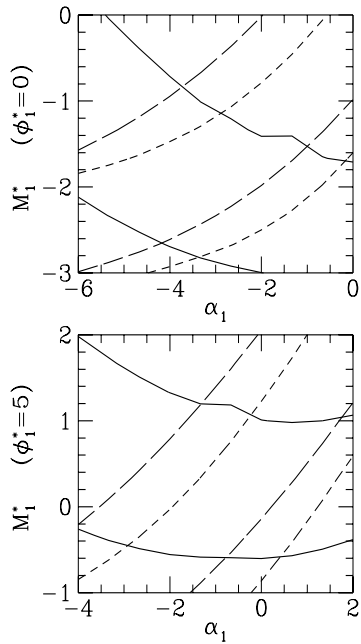

**Figure 2.** Allowed regions for different amounts of evolution as a function of $M_1^*$, $\alpha_1$ and two values of $\phi_1^*$: top $\phi_1^* = 0$ and bottom $\phi_1^* = 5$. In all cases the values of $M_0^*$ and $\alpha_0$ are allowed to vary within their uncertainties. The region within the two continuous lines corresponds to the constraint from the total number of galaxies, $N_{17-20} = 1.30 \pm 0.13 \times 10^6$. The two long-dashed lines enclose the region with mean redshift $\bar{z}_{20-21.5} = 0.22 \pm 0.01$, while the two short-dashed lines enclose the region with mean redshift $\bar{z}_{21-22.5} = 0.31 \pm 0.02$

2) The total number of galaxies in the $b_J = 17 - 20$ APM sample is:

$$N_{17-20} = 1.30 \pm 0.08 \times 10^6 . \qquad (16)$$

This number has been obtained directly from the APM maps and the error is based on a merging correction uncertainty of 5%, i.e. Maddox et al. (1990a).

3) The mean redshift at $b_J = 20 - 21.5$ found by Broadhurst et al. (1988) is:

$$\bar{z}_{20-21.5} = 0.22 \pm 0.01 . \qquad (17)$$

4) The mean redshift at $b_J = 21 - 22.5$ found by Colless et al. (1990) is:

$$\bar{z}_{21-22.5} = 0.31 \pm 0.02 . \qquad (18)$$

This list is somewhat arbitrary and does not pretend to include all observations on galaxy counts or galaxy redshifts. One could also consider other constraints, such as the slope of the galaxy counts at faint magnitudes $b_J \simeq 21$, but these are usually subject to larger uncertainties.

In Figure 2 we show the allowed regions in $M_1^*$-$\alpha_1$ space for $\phi_1^* = 0$ and $\phi_1^* = 5$. Assuming the no-evolution parameters $M_0$, $\phi_0^*$ and $\alpha_0$ in equation (15) the allowed regions for each equation (16), equation (17) and equation (18) correspond to the space between the continuous, long-dashed and short-dashed lines respectively in Figure 2. These regions correspond to 2 standard deviations (uncertainties are added in quadrature). The no evolution model (which corresponds to $M_1^* = 0$, $\alpha_1 = 0$ and $\phi_1^* = 0$) is outside the allowed region because it produces too few galaxies, but it is compatible with the mean redshift values. Models that allow only evolution in $\alpha$ (which correspond to $M_1^* = 0$ and $\phi_1^* = 0$) are not able to produce enough extra galaxies without a considerable change in the mean redshift. Models that have evolution only in $\phi^*$ (which corresponds to $M_1^* = 0$ and $\alpha_1 = 0$) require $\phi_1^* \simeq 5$ which involves a change of a factor of 2 in $\phi^*$ from $z = 0$ to $z = 0.2$.

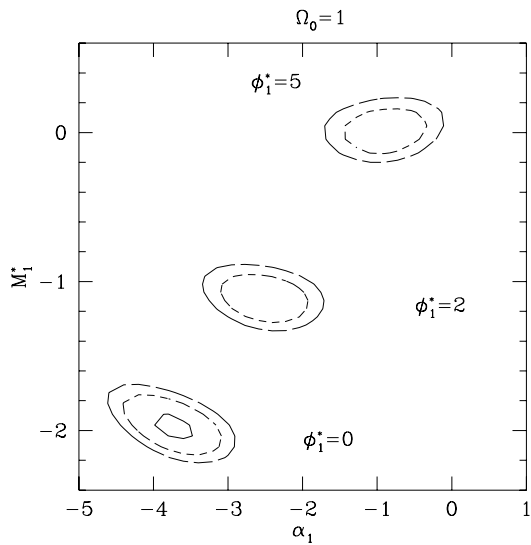

**Figure 3.** $\chi^2 = 1.32(75\%), 2.71(90\%), 3.84(95\%)$ contours for different luminosity function parameters $M_1^*$ and $\alpha_1$, and three values of $\phi_1^*$: $\phi_1^* = 0$, $\phi_1^* = 2$ and $\phi_1^* = 5$. In all cases $\Omega_0 = 1$.

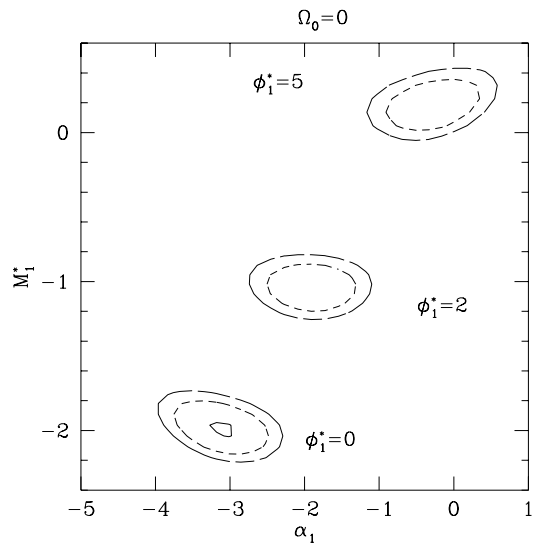

**Figure 4.** $\chi^2$ contours for different luminosity function parameters as in as Figure 3, but with $\Omega_0 = 0$

We furthermore construct a $\chi^2$ test to find contours of the values that best fit the constraints. The $\chi^2$ test has three contributions, corresponding to the differences between the predicted and observed values of $N_g$, $\overline{z}_{20-21.5}$ and $\overline{z}_{21-22.5}$.

The $\chi^2 = 1.32(75\%), 2.71(90\%), 3.84(95\%)$ contours are shown in Figure 3 for $\Omega_0 = 1$. The best fit, at a 75% level of confidence ($\chi^2 \lesssim 1.32$), corresponds to $\phi_1^* \sim 0$, $\alpha_1 \in [-4., -3.5]$ and $M_1^* \in [-2.05, -1.85]$. At a 90% level of confidence the allowed regions include other values of $\phi_1^*$. For $\phi_1^* = 5$ we have $\alpha_1 \in [-1.4, -0.3]$ and $M_1^* \in [-0.12, 0.17]$. The contours for intermediate values of $\phi_1^*$ change smoothly as $M_1^*$ increases from $\phi_1^* = 0$ to $\phi_1^* = 5$, following the trend indicated by Figure 3.

The results in Figure 3 correspond to the mean values of $M_0^*$, $\alpha_0$ and $\phi_0^*$. The uncertainties in these values can be translated roughly into further uncertainties in $M_1^*$ and $\alpha_1$ as $\Delta M_1^* \simeq \Delta M_0^*/\overline{z}$ and $\Delta\alpha_1 \simeq \Delta\alpha_0/\overline{z}$. For $\overline{z} \simeq 0.15$, in our sample, we have $\Delta M_1^* \simeq 0.9$ and $\Delta\alpha_1 \simeq 1$, comparable with the values shown by the contours.

The above results depend slightly on the value of $\Omega_0$. As we change $\Omega_0$ from 1 to 0 the contours move slowly to the right, with mean values $\alpha_1$ that follow $\alpha_1(\Omega_0) \sim \alpha_1(0) - \Omega_0$. There is also a smaller trend upwards, which is correlated with changes in $\phi_1^*$, so that mean values of $M_1^*$ follow $M_1^*(\Omega_0) \sim M_1^*(0) - \Omega_0\phi_1^*/20$. The results for $\Omega_0 = 0$ are shown in Figure 4.

We find $\chi^2 \simeq 15.6$ ($\Omega_0 = 0$) or $\chi^2 \simeq 28$ ($\Omega_0 = 1$) for the luminosity function proposed by Maddox et al. (1990a), i.e. in our notation $M_0^* = -19.8$, $\alpha_0 = -1$, $M_1^* = -2$ and $\alpha_1 = -2$ normalized using $\phi_0^*$ from equation (15). This model has been used by Maddox et al. and by G94 as a rough approximation to the selection function when estimating the three-dimensional clustering in the APM. The reason for the large $\chi^2$ is that this model predicts larger mean redshifts than those observed: $\overline{z}_{20-21.5} \simeq 0.26$ for $\Omega_0 = 0$ or $\Omega_0 = 1$ and $\overline{z}_{21-22.5} \simeq 0.34(\Omega_0 = 1)$ or $\overline{z}_{21-22.5} \simeq 0.33(\Omega_0 = 0)$. With such a large $\chi^2$ it is clear that this model does not fit well the constraints we have chosen and it is important to see how different the resulting clustering predictions are.

Baugh & Efstathiou (1993) have proposed a functional form for the redshift distribution $N(z)$ which provides an acceptable match to the deep redshift histograms (Broadhurst et al. 1988 and Colless et al. 1990) whilst simultaneously fitting the Stromlo/APM redshift distribution (Loveday et al. 1992). By construction one can also match the total number in the $b_J = 17 - 20$ APM sample. We find that this $N(z)$ distribution gives $\overline{z}_{20-21.5} \simeq 0.23$ and $\overline{z}_{21-22.5} \simeq 0.29$, and this contributes to giving an acceptable $\chi^2 \simeq 2.23$. Nevertheless, when using this functional form for $N(z)$ we do not explore the uncertainty in the selection function, which is what we need to do in order to study its effect on the projected clustering.

## 4  EVOLUTION OF CLUSTERING

As usual, we parametrize the evolution of clustering using:

$$\xi_2(r, z) = (1 + z)^{-(\epsilon+3)} \xi_2(r), \tag{19}$$

where $\xi_2(r)$ corresponds to $z = 0$. For the higher order correlations we assume that the evolution follows from the hierarchical model equation (2). That is, that $Q_J$ and $S_J$ do not evolve much with $z$. This is in good agreement with N-body simulations (see Baugh, Gaztañaga & Efstathiou 1995 and Gaztañaga & Baugh 1995).

When the intrinsic clustering properties do not evolve in proper coordinates (stable clustering) the excess probability of finding a galaxy at separation $r$ from a given galaxy is a constant:

$$n(z)\ \xi_2(r,z) = n_0\ \xi_2(r) = const, \tag{20}$$

where $n_0$ corresponds to $n(z)$ at $z = 0$. Because of the Universal expansion $n(z) = n_0(1 + z)^3$, so that:

$$\xi_2(r,z) = (1+z)^{-3} \xi_2(r), \tag{21}$$

corresponding to $\epsilon = 0$ in equation (19). Thus $\xi_2(r,z)$ is smaller by a factor of $\sim 1.7$ at $z \sim 0.02$.

However, clustering evolves under gravity. When $\xi_2 < 1$, i.e. at large scales $r > r_0$, the clustering is near the linear regime $\xi_2(x) = a^2 \xi_2^0(x)$, where $\xi_2^0(x)$ are the initial conditions. Assuming that over some range in scale the initial conditions are scale invariant: $\xi_2^0(x) \sim x^{-\gamma}$, we have $\epsilon = \gamma - 1$ in equation (19). ¿From observational galaxy catalogs (see below) it is found that $\gamma \simeq 2$ for scales $r \gtrsim r_0$ increasing to larger values at larger scales, so that $\epsilon \gtrsim 1$ or larger. Thus in the linear regime $\epsilon \gtrsim 1$ and $\xi_2(r,z)$ evolves more than in the stable clustering regime. In the small scale regime, where $\xi_2 \gtrsim 1$, N-body simulations typically show more clustering evolution than in the linear regime (i.e. Baugh, Gaztañaga & Efstathiou 1995, see also Figures 11-12), and therefore $\epsilon > \gamma - 1$. Note nevertheless that these arguments apply to the matter distribution, while we are interested in the galaxy distribution.

If the clustering pattern is fixed in comoving coordinates, then $\epsilon = \gamma - 3 \simeq -1.3$, with less evolution than in the stable clustering regime. This might describe some models in which galaxies are identified with high density peaks. As one would expect peaks to move less than mass particles it results in less evolution for the clustering of galaxies.

Although there are several observations of clustering of faint galaxies they do not seem to set a definitive constraint on $\epsilon$. We will consider the range $\epsilon \simeq -1.3$ to $\epsilon \simeq 1.3$ below, in order to include all the above possibilities.

## 5   CLUSTERING IN THE APM SURVEY

We use here the clustering results from the angular APM Galaxy Survey (Maddox *et al.* 1990a), in particular for the $b_J = 17 - 20$ sample which has over $1.3 \times 10^6$ galaxies. The area-averaged angular J-point correlation functions $\overline{w}_J(\theta)$ were estimated for $J = 2 - 9$ by G94. We have recalculated here these correlations $\overline{w}_J(\theta)$ using more cell sizes so that we have better scale resolution (each cell is 50%, instead 100 %, larger than the previous one).

### 5.1   The Variance: $\overline{\xi}_2$

#### 5.1.1   *The 2-point amplitude: $r_0$ and $\sigma_8$*

Figure 5 shows the estimated values of $\overline{w}_2$. This figure is similar to Figure 1 of G94, but with more cell sizes and with errors from the variance in 4 random subsets (and not 4 zones). The possible effects of any artificial gradients in the APM caused by plate matching errors could introduce the

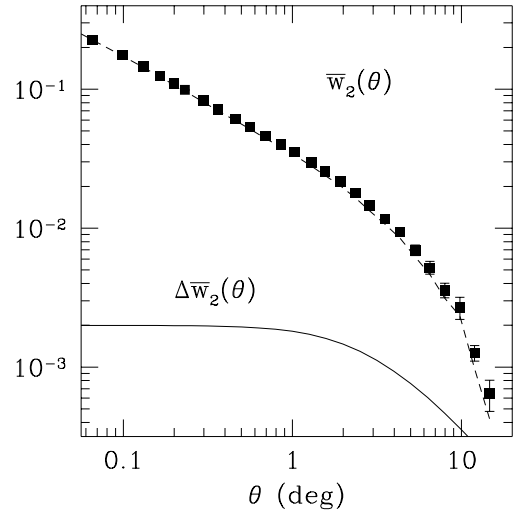

**Figure 5.** Variance of angular counts, $w_2(\theta)$, for the $b_J = 17-20$ APM sample. Errors are estimated from the variance in 4 random subsets. The continuous line models the effect of a possible artificial gradient in the APM Survey caused by plate matching errors (derived using the model of Maddox *et al.* 1990c). The dashed line shows the result of substracting this correction from our mean estimate.

following contribution to the two-point angular correlation function:

$$\Delta w_2(\theta_{12}) \lesssim 2 \times 10^{-3} / [1 + (\theta_{12}/6°)^2]. \tag{22}$$

This function varies from $2 \times 10^{-3}$ at small angular scales to $1 \times 10^{-3}$ on the plate scale $\theta_{12} \simeq 6°$ in the model for large scale gradients described by Maddox *et al.* (1990c, see also Baugh & Efstathiou 1993). The corresponding angular variance $\Delta \overline{w}_2$ in circular cells of radius $\theta$ is shown as a continuous line in Figure 5. The possible net effect on the measured $\overline{w}_2$ is shown as a dashed line. The resulting angular variance lies within the estimated sampling errors.[*] Other uncertainties in the construction of the APM Galaxy Survey (such as other magnitude errors or non-uniform star-galaxy separation) are likely to give even smaller errors (Maddox *et al.*, 1990c).

A least-squares fit to a power-law $\overline{w}(\theta) = A\ \theta^{-\beta}$ for small scales, $\theta = 0.09 - 0.9$ in degrees, gives $\beta \simeq 0.7 \pm 0.2$ and $A = (3.82 \pm 0.12) \times 10^{-2}$, where we have applied a merging correction of 5 %, (see Maddox *et al.* (1990a)). [Note that this correction has not been applied to the values shown in Figure 5.] This value of $A$ is used in equation (7) to find $r_0$. Given a model for clustering evolution, i.e. $\epsilon$ in equation (19), the value of $r_0$ depends on each parameter $\alpha_1$, $M_1^*$ and $\phi_1^*$ in $\phi(L,z)$, ( equation (14)). In Figure 6 we show these values of $r_0$ as contours in the $\alpha_1$-$M_1^*$ plane for different $\epsilon$ and two values of $\Omega_0$. The allowed contours from Figure 3 are placed on top of the $r_0$ values.

---

[*] Note that at large scales, where this effect is larger, our final errors in $\overline{\xi}_2$ are dominated by the uncertainty in the value of the local slope $\gamma$. These final errors are much larger than the artificial gradient correction, see Figures 7-8.

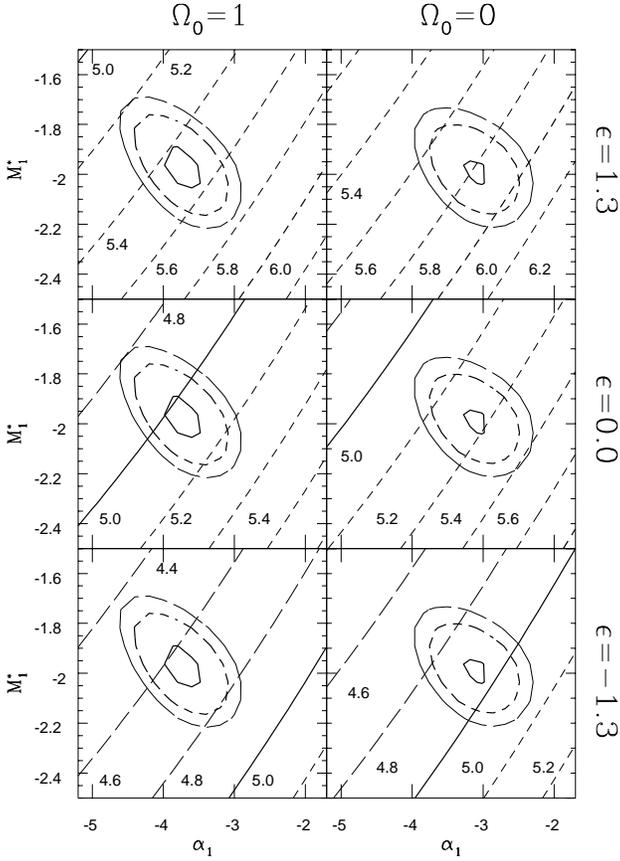

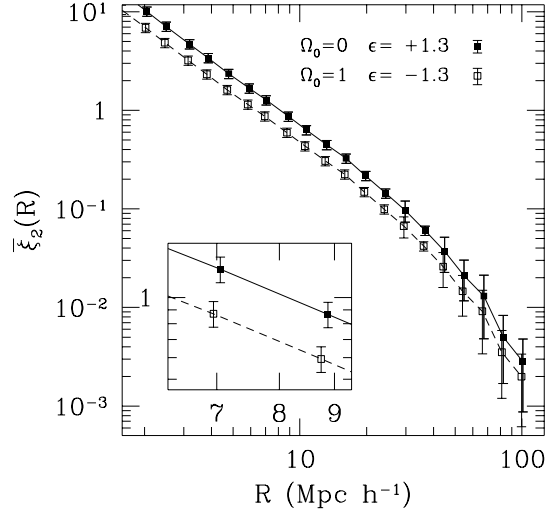

**Figure 7.** Comparison of different estimates of $\bar{\xi}_2$ from direct inversion of the angular variance $\bar{w}_2(\theta)$ for the two extreme situations: i) $\Omega = 1$ and $\epsilon = -1.3$ (open squares) ii) $\Omega = 0$ and $\epsilon = +1.3$ (filled squares). In both cases $\phi_1^* = 0$ and $M_1^* = -2$. The inset is an enlarged view of a portion of the same graph around $R \simeq 8\,h^{-1}$ Mpc.

**Figure 6.** Contours of values of $r_0$ for the APM $b_J = 17 - 20$ galaxy sample using different luminosity functions, i.e. values of $\alpha_1$ and $M_1^*$, and $\phi_1^* = 0$. The thick continuous line marks the contour where $r_0 = 5\,h^{-1}$ Mpc, above this line $r_0 < 5\,h^{-1}$ Mpc. Other boundaries are marked at intervals of $0.2\,h^{-1}$ Mpc (long-dashed lines correspond to $r_0 < 5\,h^{-1}$ Mpc, short-dashed lines to $r_0 > 5\,h^{-1}$ Mpc ). Each panel shows a different model for the evolution of clustering labelled by $\epsilon$ and $\Omega$. Overlaid closed contours show the allowed regions according to the number counts and mean redshift, i.e. Figure 3.

**Table 1.** Estimated values of $\sigma_8$ in the APM for different models for the evolution of clustering and different values of $\Omega_0$.

| $\sigma_8^{APM}$ | $\Omega_0 = 1$ | $\Omega_0 = 0$ |
|---|---|---|
| $\epsilon = +1.3$ | $0.95 \pm 0.07$ | $1.00 \pm 0.08$ |
| $\epsilon = 0.0$ | $0.89 \pm 0.06$ | $0.93 \pm 0.07$ |
| $\epsilon = -1.3$ | $0.83 \pm 0.05$ | $0.87 \pm 0.06$ |
| $\epsilon = -3.0$ | $0.75 \pm 0.05$ | $0.79 \pm 0.06$ |

Figure 6 corresponds to the case $\phi_1^* = 0$, but overall, for any given $\epsilon$ the final allowed values of $r_0$ are roughly independent of $\phi_1^*$.

It is also interesting to express the 2-point amplitude in terms of the variance $\sigma_8^2$ in spheres of radius $8\,h^{-1}$ Mpc, i.e. equation (4). In Table 1 we show the best fit values of $\sigma_8$.

For each value of $\epsilon$ and $\Omega_0$ we find the most likely value of $r_0$ and its uncertainty for *all* possible values of $M_1^*$, $\alpha_1$ and $\phi_1^*$. The resulting uncertainties are then added in quadrature with the sampling errors in $A$ ($\simeq 2\%$) to estimate $\sigma_8$ from equation (4) –we use the slope at $1°$, $\gamma = 1.70 \pm 0.02$, which corresponds to $R \simeq 8\,h^{-1}$ Mpc. Thus the resulting errors in Table 1 include sampling errors, uncertainties in the slope, and uncertainties in the selection function. This adds up to an $\simeq 8\%$ error in $\sigma_8$, for any given value of $\epsilon$ and $\Omega_0$. There is 7% variation of $\sigma_8$ with $\epsilon$ (for $\epsilon$ between -1.3 and 1.3) and a 3% variation with $\Omega_0$.

In Table 1 we also show the results for $\epsilon = -3$, which corresponds to the case of no clustering evolution and no cosmic expansion.

### 5.1.2 *The shape of* $\bar{\xi}_2$

We next use the inversion presented in §2 to estimate $\bar{\xi}_2$ at different scales.

Figure 7 shows the extreme case $\epsilon = +1.3$ and $\Omega = 0$ as compared to $\epsilon = -1.3$ and $\Omega = 1$. In both cases, we use the luminosity function model with $\phi_1^* = 0$ and $M_1^* = -2$, with $\alpha_1 = -4$ for $\Omega = 1$ and $\alpha_1 = -3$ for $\Omega = 0$, suggested by figures 3-4. Results for other models for the luminosity function or evolution of clustering give intermediate values (see Figure 6). The inset shows the amplitude of $\bar{\xi}_2$ around $R \simeq 8\,h^{-1}$ Mpc, i.e. $\sigma_8^2$, which is in good agreement with the values given above.

Figure 8 compares the values of $\bar{\xi}_2(R)$ for the model with $\alpha_1 = -4$ with the ones for the model proposed by Maddox *et al.* 1990a, with $\alpha_1 = -2$. The net effect is a higher amplitude for the latter model, with $\sigma_8^2 \simeq 1.0$ instead of $\sigma_8^2 \simeq 0.8$. In both cases $\phi_1^* = 0$, $M_1^* = -2$, $\epsilon = 0$ and $\Omega_1 = 1$.

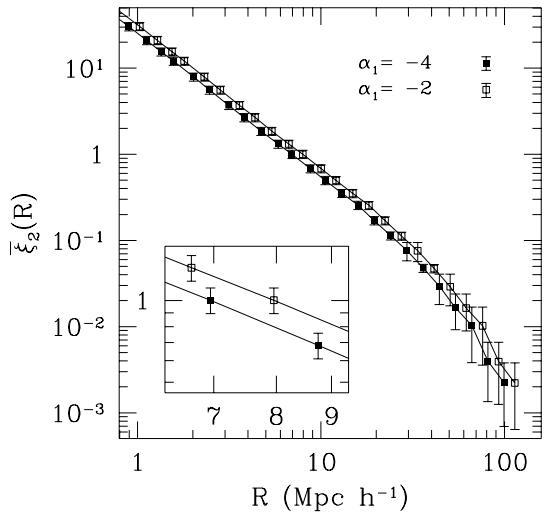

**Figure 8.** Comparison of two different estimations of $\overline{\xi}_2$ from direct inversion of the angular variance $\overline{\omega}_2(\theta)$ using different selection functions ($\epsilon = 0$). Filled squares correspond to $\alpha_1 = -4$, while open squares correspond to the model by Maddox *et al.* 1990a, which has $\alpha_1 = -2$. The inset is an enlarged view of a portion of the same graph around $R \simeq 8\, h^{-1}$ Mpc.

¿From Figures 7-8 we can see that the amplitude of $\overline{\xi}_2(R)$, and not the shape, is affected by the uncertainties in the selection function.

## 5.2 Higher order correlations $\overline{\xi}_J$

### 5.2.1 The amplitudes $r_J$

We next find the projection coefficients $r_J$ using equation (7). In Figures 9 we show the values of $r_3$ as contours in the $\alpha_1$-$M_1^*$ plane for different values of $\epsilon$ and $\phi_1^* = 0$. We find similar results for other values of $\phi_1^*$.

The variation of $r_3$ for different models of the selection function, (i.e. in the $\alpha_1$-$M_1^*$ plane) is quite small, typically $\Delta r_3 \simeq 0.01$ or smaller within the allowed regions. This variation is smaller than that involving the 2-point function amplitudes, e.g. $r_0$. This is because $r_3$ corresponds to a relative amplitude, i.e. $\xi_3/\xi_2^2$, and therefore there is some degree of cancelation between the values of $I_k$ (note that $r_J$ are dimensionless in powers of $I_k$ while $r_0$ is not).

The variation for different clustering models, i.e. as a function of $\epsilon$, is more significant. For $\epsilon = 0$ we have $r_3 \simeq 1.19$ ($r_3 \simeq 1.20$) for $\Omega_0 = 1$ ($\Omega_0 = 0$), while $\epsilon = -1.3$ gives $r_3 \simeq 1.169$ ($r_3 \simeq 1.176$) and for $\epsilon = 1.3$, $r_3 \simeq 1.217$ ($r_3 \simeq 1.226$). Thus the overall range is quite large, $r_3 \simeq 1.17 - 1.23$, but still only represents a 5% variation compared to the 32% variation in $r_0$ in Figure 6.

The general pattern is similar for higher orders with the overall variation increasing to about 10%, 15%, 21%, 28%, 34% and 41% for $J = 4, 5, 6, 7, 8$ and 9, respectively.

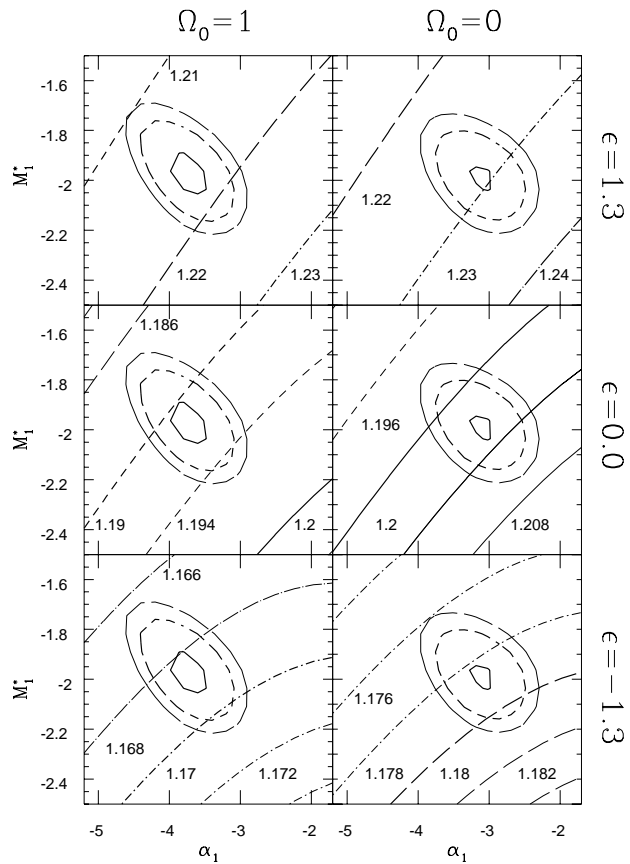

**Figure 9.** Contours of values of $r_3$ for the APM $b_J = 17 - 20$ galaxy sample using different luminosity functions, i.e. values of $\alpha_1$, $M_1^*$, and $\phi_1^* = 0$. Each line marks the boundary for contours of $r_3$ values, as marked. Each panel shows a different model for the evolution of clustering labeled by $\epsilon$ and $\Omega$ ($\Omega = 1$ except for the top right-hand corner, which has $\Omega = 0.2$). The contours show the allowed regions according to the number counts and mean redshift of Figure 3.

### 5.3 The shape of $S_J$

We next use the inversion presented in §2 to estimate $S_J$ from $s_J$ at different scales. Figure 10 compares two extreme cases: i) $\Omega = 1$ and $\epsilon = -1.3$ (open squares) ii) $\Omega = 0$ and $\epsilon = +1.3$ (filled squares). In both cases $\phi_1^* = 0$ and $M_1^* = -2$. As pointed out above, the net effect of changing the selection function is small and the differences in Figure 10 are dominated by the changes in $\epsilon$.

## 6 DISCUSSION

### 6.1 The value of $\sigma_8$

Our final set of selection functions differ from that proposed by Maddox *et al.* (1990a) for inverting the APM 2-point correlation, and also used by Gaztañaga (1994) to estimate higher order correlations. The difference, though small, is significant given the errors. The net effect is a higher amplitude for the selection function proposed by Maddox *et al.*, with $\sigma_8^2 \simeq 1.0$ for $\epsilon = 0$, instead of $\sigma_8^2 \simeq 0.8$ (see Figure 7).

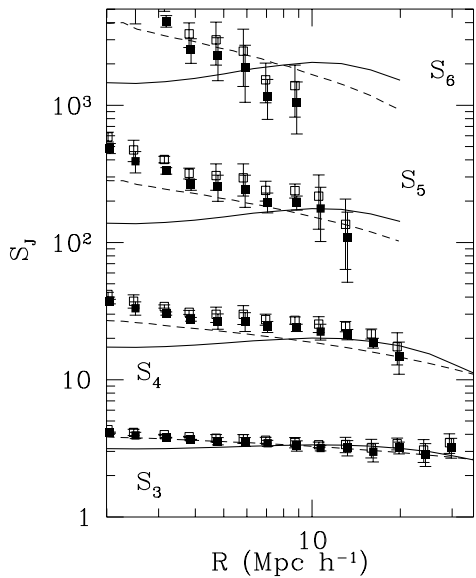

**Figure 10.** Comparison of two different estimates of $S_J(R)$ from direct inversion of the angular variance $s_J(\theta)$ using different selection functions. Points show the two most extreme cases, as in Figure 7. Lines correspond to the perturbation theory predictions for the linear power spectrum $P(K)$ of the $\Gamma = 0.2$ CDM model (solid line), and the estimated $P(k)$ in the APM (Baugh & Efstathiou 1993) (dashed line).

The effect on the hierarchical amplitudes $S_J$ for the higher order correlations is smaller. On the other hand, our best models for the selection function (cf Figure 3) agree well with the functional form for the redshift distribution $N(z)$ proposed by Baugh & Efstathiou (1993). However, the latter does not explore the uncertainties in the selection function.

If galaxy clustering in the APM grows according to the gravitational growth of matter fluctuations ($\epsilon \simeq 1.3$), then $\sigma_8^{APM} = 0.95 \pm 0.07$ ($1.00 \pm 0.08$) for $\Omega_0 \simeq 1$ ($\Omega_0 \simeq 0$), close to the values measured from nearby optical samples such as the North Zwicky Center for Astrophysics catalog (Huchra et al. 1983, hereafter CfA) or the Southern Sky Redshift Survey (Da Costa et al. 1991, hereafter SSRS). On the other hand, if we assume that clustering evolution is fixed in comoving coordinates ($\epsilon \simeq -1.3$), we find $\sigma_8^{APM} = 0.83 \pm 0.05$ ($0.87 \pm 0.06$), closer to the value for IRAS galaxies (e.g. Fisher et al. 1994). On comparing with redshift samples, one should take into account the possible effect of peculiar velocities. Fry & Gaztañaga (1994) have used configurations that minimize redshift distortions to estimate the clustering in real space in the CfA, SSRS and 1.9 Jy IRAS catalogues. For our comparison we focus on their results for volume limited samples CfAN80, SSRS80 and IRAS65, which represent a compromise between sampling a large volume and having a large enough galaxy density. The clustering in the smaller nearby samples is probably affected by large scale density fluctuations. We use the quoted mean values of $r_0$ and $\gamma$ and its uncertainties from Table 4 of Fry & Gaztañaga (1994) to find $\sigma_8$ from equation (4).

The resulting values for $\sigma_8$ are shown in Table 2. These values are compatible with the estimates made by Fisher et al. (1994) for the 1.2 Jy IRAS and those for the CfA (from Davis & Peebles 1983): $\sigma_8^{IRAS} = 0.69 \pm 0.04$ and $\sigma_8^{CfA} = 0.95 \pm 0.06$, but we find larger errors. This could be partially due to differences in the method of estimation. The variance coming from the finite size of the sample (i.e. caused by fluctuations on the scale of the sample) could be quite important for these nearby catalogues. Fry & Gaztañaga (1994) do include a finite volume contribution to the error by modelling the tail of the probability distribution. In fact, the difference between the estimates of $\sigma_8$ from the CfA and SSRS samples in Table 2 and also the difference between the two estimates mentioned for $\sigma_8^{CfA}$ (which use different subsets of the CfA galaxies), indicate that the effect of the combination of finite volume and (possible) differential selection biases for optical galaxy samples is as large as 25%. This is in good agreement with the error estimates in Table 2.

**Table 2.** Values of $\sigma_8$ in real space estimated for CfA, SSRS and IRAS samples volume limited to a depth $\mathcal{D}$, and covering a total volume $V_T$ in $(h^{-1}\,\mathrm{Mpc})^3$.

| Sample | $\mathcal{D}$ | $V_T$ | $\sigma_8$ |
|---|---|---|---|
| CfA | $80\,h^{-1}$ Mpc | $3.0 \times 10^5$ | $1.21 \pm 0.21$ |
| SSRS | $80\,h^{-1}$ Mpc | $2.9 \times 10^5$ | $0.97 \pm 0.25$ |
| IRAS | $65\,h^{-1}$ Mpc | $1.0 \times 10^6$ | $0.70 \pm 0.21$ |

The precise values of $\sigma_8$ from the APM galaxies provide important constraints on models for structure formation. For example, there is a whole range of models for which the predicted values of $\sigma_8$ for matter fluctuations in linear theory, $\sigma_8^L$, turn out to be $\sigma_8^L \geq 1$ when normalized to COBE fluctuations (Gorski, Stompor & Banday 1995, Stompor, Gorski & Banday 1995, Gorski et al. 1995). Given that the non-linear values of clustering are typically similar or larger than the linear values at $8\,h^{-1}$ Mpc (see Figure 11-12), we see that the amplitude of matter fluctuations in these models are larger than the fluctuations estimated from APM galaxies in Table 1. Thus these models require that galaxies are anti-biased with respect to the mass, i.e. that galaxy formation (and/or galaxy selection) is less likely to occur in large scale high density peaks and more likely to occur in large scale low density voids. Although not impossible, this would require a peculiar selection effect which needs to be studied.

In Figures 11 and 12 we compare the APM variance $\overline{\xi}_2^{APM}$ (open squares) for the case with highest amplitude, i.e. $\epsilon = 1.3$, with the linear (continuous line) and non-linear (closed squares) predictions for matter fluctuations in the $\Gamma = 0.2$ ($\Omega_0 = 0.2$, $\lambda = 0.8$, $h = 1$) CDM model and the $\Gamma = 0.5$ ($\Omega_0 = 1$, $h = 0.5$) CDM model (from Gaztañaga & Baugh 1995). Both models are normalized to $\sigma_8^L = 1$, which corresponds to the lower end of the COBE normalization even when a small tensor contribution is allowed. The inset shows the bias factor $b^2 \equiv \overline{\xi}_2^{APM}/\overline{\xi}_2^{CDM}$ as a function of scale, where $\overline{\xi}_2^{CDM}$ is either the linear (continuous line) or non-linear matter variance (closed squares). For a local biasing transformation, one would expect that at large scales, where $\overline{\xi}_2 < 1$, there should be a linear relation between galaxy and matter $\overline{\xi}_2$, independent of scale (see Fry & Gaztañaga 1993). This does not seem to happen in the "standard" CDM model (Figure 12), which is the reason why this model is not favored by the clustering of APM galaxies (Maddox et al. 1990a). The open model (Figure 11) shows

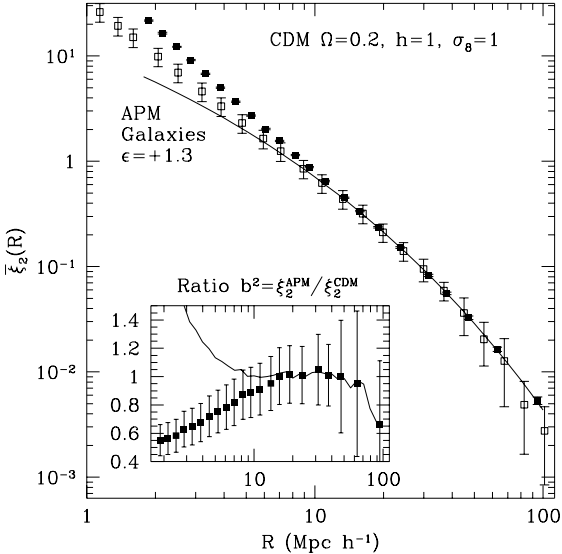
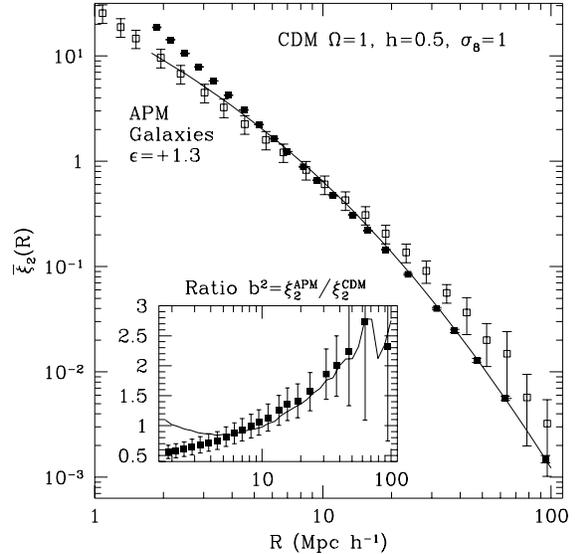

**Figure 11.** Comparison of the APM galaxy variance $\overline{\xi}_2^{APM}$ (open squares) with the variance in the matter fluctuations of the $\Gamma = 0.2$ CDM model ($\Omega = 0.2$, $h = 1$, $\sigma_8 = 1$) measured from an N-body simulation (filled squares). The continuous line corresponds to the linear theory prediction for the same model. The projection effects in the APM have been modelled so as to give the highest amplitude. The inset shows the ratios of the galaxy variance to the matter variance $b^2 = \overline{\xi}_2^{APM}/\overline{\xi}_2^{CDM}$ (on a linear scale).

**Figure 12.** Comparison of $\overline{\xi}_2(r)$ values in the APM survey and for the $\Gamma = 0.5$ CDM model ($\Omega = 1$, $h = 0.5$, $\sigma_8 = 1$). The format and symbols are the same as in Figure 11.

a constant $b^2$ for $R \geq 10\,h^{-1}$ Mpc, but its value is smaller than one at all scales (anti-bias).[†]

### 6.2 Higher order correlations

The final uncertainty in the range of values of $S_J$ is typically twice the estimated sampling errors, with the highest values for the case of less clustering evolution (Figure 10). Again, the precise values of $S_J$ from the APM could provide important constraint on models for structure formation. If galaxy fluctuations trace matter fluctuations, one can predict the values of $S_J$ in the APM at large scales by just assuming gravitational growth from initially small gaussian fluctuations. These predictions depend on the shape (but not the amplitude) of the initial power spectrum, and are only valid at large scales $\overline{\xi}_2 \lesssim 1$ or $R \gtrsim 10\,h^{-1}$ Mpc (Fry 1984b, Juszkiewicz, Bouchet & Colombi 1993, Bernardeau 1994, Juszkiewicz et al. 1995, Lokas et al. 1995, Gaztañaga & Baugh 1995, Baugh, Gaztañaga & Efstathiou 1995). They are shown as continuous lines (APM final power-spectrum) and dashed lines (low $\Omega$ CDM initial power-spectrum) in Figure 10 (from Gaztañaga & Frieman 1994).

Bernardeau (1995) has critized the use of the "tree" hierarchy (2) to model the projection effects in angular catalogues because the PT results are not exact tree models with constant values of $Q_J$. Bernardeau favours using the 2D projected PT predictions directly in comparisons with the angular data. The problem with this approach is that the PT results are only valid on large scales while the projection mixes small with large scales. ¿From the N-body results mentioned above it is clear that, at least at scales where $\overline{\xi}_2 > 1$, the PT hierarchy has the wrong amplitudes. This could introduce spurious projection effects, which are particularly bad for the PT hierarchy as it is not scale invariant. Moreover, the small angle approximation $\theta \to 0$ has a different meaning when the hierarchy is not scale invariant, which could explain why the Bernardeau results seem to be so sensitive to this approximation. In contrast, the quasi-scale invariant approach presented here (and used in G94) is optimal in the sense that it allows the hierarchy to be scale dependent (as in the PT case) but without introducing the spurious amplitudes at small scales. Our preliminary tests of projections in N-body simulations seem to verify the validity of the quasi-scale invariant model.

### 7 CONCLUSION

We have studied the uncertainties involved in the estimation of three dimensional clustering properties from the angular distribution in a sample $b_J = 17 - 20$ from the APM Galaxy Survey. We have considered the effect of a change in the selection function and the changing of clustering amplitudes with redshift.

The final range of values for the inverted 2-point amplitude, covering all possibilities, is quite large, $\sigma_8 \simeq 0.78 - 1.08$ (see Table 1). The predicted shape for $\overline{\xi}_2$ is not much affected by any of these uncertainties (see Figures 7-8). The values of $\sigma_8$ in the nearby samples, Table 2, can be compared directly with those in Table 1 for the APM Survey. We can see that the sampling and projection errors in $\sigma_8$ for the nearby

---

[†] Although $b^2$ is close to unity at large scales in Figure 11, the mean COBE normalization for this model *without* any tensor contribution is $\sigma_8^L \simeq 1.3$, i.e. $\overline{\xi}_2^{CDM}$ should be scaled up a factor $\simeq 1.7$ producing $b^2 \simeq 0.5$

catalogues are two to three times larger than those in the APM, but the uncertainties in the cosmological parameters and the evolution of clustering are less important.

In a recent preprint, Loveday *et al.* (1995) have estimated $\sigma_8^2 = 0.90 \pm 0.05$ for the real space variance in the Stromlo/APM Redshift Survey, i.e. $\sigma_8 = 0.95 \pm 0.03$. A comparison with the values in Table 1 suggests a large value of $\epsilon$, i.e. as expected if clustering grows according to gravity in an expanding universe with little or no biasing. In particular, the value of $\sigma_8$ in the Stromlo/APM is clearly incompatible with the one in Table 1 for the whole APM if $\epsilon = -3$, indicating that both the Universal expansion and some degree of clustering evolution are neccessary to reconcile the two measurements of $\sigma_8$.

The results for higher order correlations in Figure 10 agree with the results in Figure 2 of Gaztañaga & Frieman 1994, where it was pointed out that the observed values of $S_J$ in the APM Survey are compatible with the clustering that emerges from gravitational growth of small (initially Gaussian) fluctuations, regardless of the cosmological model we assume for the universe, i.e. $\Omega$, $\lambda$, $H_0$ or the nature of dark matter. Again we find here that the observations requiere little or no biasing to match the gravitational predictions.

Some of the uncertainties considered in our analysis could be removed with a better knowledge of the redshift distribution $N(z)$ for galaxies in the $b_J = 17 - 20$ range, while a better understanding of galaxy formation seems necessary to make more detailed predictions about galaxy clustering evolution and the possibility of *bias* or *anti-bias* in the final galaxy distribution.

## Acknowledgements

I thank Rupert Croft, Cedric Lacey and Michael Strauss for their useful comments on the manuscript. I also want to thank the Astrophysics group in Oxford for their hospitality. This work was supported in part by a Fellowship from the Commission of the European Communities and from Consejo Superior de Investigaciones Científicas, CSIC, in Spain.